# Domain wall pinning in FeCoCu bamboo-like nanowires

*Eider Berganza, Cristina Bran\*, Manuel Vázquez and Agustina Asenjo*

E. Berganza, Dr. C. Bran, Prof. M. Vázquez, Dr. A. Asenjo
Institute of Materials Science of Madrid, CSIC, 28049 Madrid, Spain

The pinning effect of the periodic diameter modulations on the domain wall propagation in FeCoCu individual nanowires is determined by Magnetic Force Microscopy, MFM. A main bistable magnetic configuration is firstly concluded from MFM images characterized by the spin reversal between two nearly single domain states with opposite axial magnetization. Complementary micromagnetic simulations confirm a vortex mediated magnetization reversal process. A refined MFM imaging procedure under variable applied field allows us to observe metastable magnetic states where the propagating domain wall is pinned at certain positions with enlarged diameter. Moreover, it is demonstrated that in some atypical nanowires with higher coercive field it is possible to control the position of the pinned domain walls by an external magnetic field.

1. Introduction

Novel phenomena as spin-transfer torque, Dzialoshinskii-Moriya interaction or magnetic skyrmions are currently intensively investigated in thin-film and planar nanostructures for modern spintronics devices[1,2,3]. Particularly, the control over the magnetic domain wall (DW) motion is a key aspect for the development of spintronics, logic systems or sensing devices[4,5,6]. A large number of works are currently devoted to the study of DW dynamics along ferromagnetic elements driven by electric current[7,8,9] or magnetic fields[10,11]. In order to develop DW-based applications, the control and understanding of the DW configuration and its pinning/depinning mechanism becomes essential. The use of well-localized artificial pinning sites enables the trapping of DWs at given positions ensuring the needed high thermal stability and the reproducibility of the DW motion. The most widespread method of creating pinning centers in nanostripes is patterning notches with different shapes in planar nanostructures[12,13,14]. However, much less attention has been paid to DWs in cylindrical nanowires and to their pinning mechanisms either from the theoretical or experimental viewpoints [15,16,17]. It is worth noticing that this geometry favors the development of vortex or hybrid domain walls instead of transverse domain walls[18,19]. In contrast to the two dimensional structures, vortex domain walls in cylindrical nanowires move uniformly which in turn should be more convenient for applications[11]. Even more, field driven vortex domain walls in cylindrical nanowires do not show a Walker breakdown[20].

Ferromagnetic cylindrical nanowires with negligible crystalline anisotropy and very high aspect ratio have been investigated. Their longitudinal uniaxial magnetic anisotropy makes them ideal systems to study the magnetization reversal process. A significant work has been devoted to the preparation of ferromagnetic cylindrical nanowires grown into templates by electrochemical route[21,22]. A first main goal of the study has been the growth of cylindrical nanowires with periodical small segments of different diameter that we label as bamboo-like structure. CoFe based alloy nanowires have been selected due to their large saturation magnetization and high Curie temperature, which makes them good candidates to replace rare-earth free based permanent magnets in certain applications.

The final aim of this study is to determine the local magnetic configuration along individual nanowires and to show the pining effect of the bamboo-like geometry in FeCoCu nanowires making use of advanced Magnetic Force Microscopy (MFM) technique[23,24]. MFM is a recognized powerful technique to image the local spin configuration[25] as well as the magnetization reversal process at the nanoscale[26]. This technique provides high resolution images of the magnetic configuration (around 20nm) together with the corresponding topographic information. MFM images have been firstly obtained at various remanence states and under an external applied magnetic field parallel to the cylinder axis. To gain deeper understanding of the magnetization reversal process, a more subtle imaging procedure has been used. Finally, complementary micromagnetic simulations were carried out using object orientated micromagnetic framework[27] (OOMMF) package to confirm the experimental results.

2. **Bamboo-like nanowires**

Nanowires with very high aspect ratio were grown by electrodeposition into the pores of anodic alumina membranes. Modulated pores are produced by pulsed hard anodization in oxalic aqueous solution. By tunning the electrochemical parameters we could introduce periodical changes into the diameter in a controlled way.

The total length of the nanowires is about 12 μm. They display a bamboo-like structure with a periodicity of 800nm and with diameters of about 150nm (segment) and 170nm (modulation), respectively (See **Figure 1**). The composition, determined by EDS, of the nanowires alloy is $Fe_{28}Co_{67}Cu_5$ (hereafter referred to as the FeCoCu nanowires).

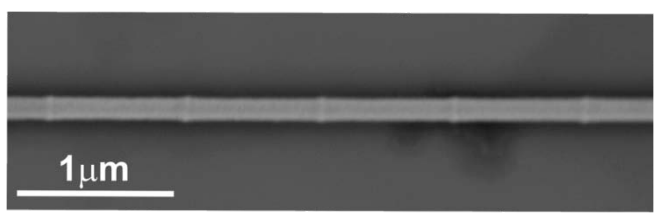

**Figure 1.** Scanning electron microscopy image of an individual nanowire.

3. **Magnetic Force Microscopy imaging in remanence**

Nanowires of constant diameter (φ= 150nm) and similar length were also measured as a reference to evaluate the effect of the diameter modulation. MFM images were obtained in the demagnetized state.

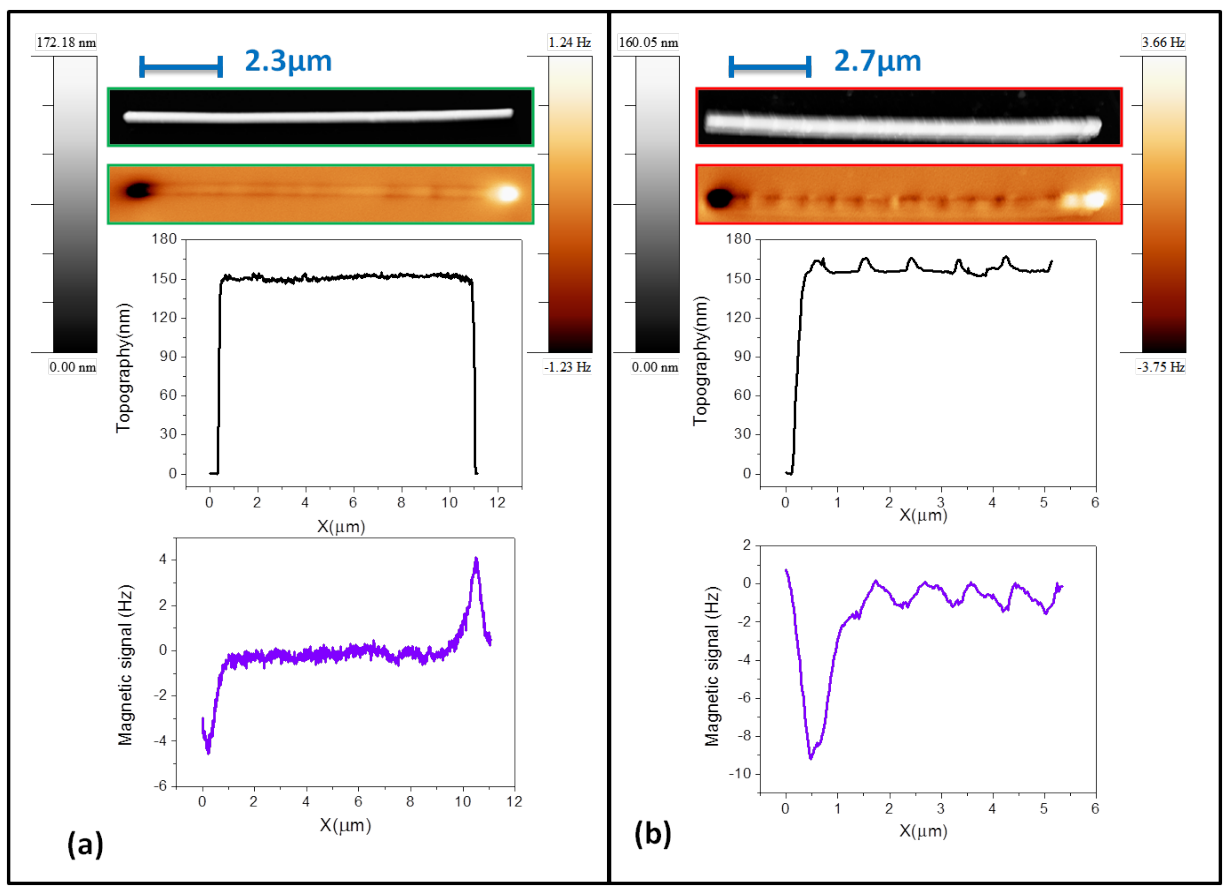

**Figure 2.** Topographic and MFM images of (a) a NW of homogeneous diameter showing a single domain configuration and (b) a bamboo-like NW where the local modulations give rise to small divergences of the magnetization. Profiles along the main NWs axis were performed in both (only left part of the bamboo-like NW is shown for clarity purposes).

A significant number of, homogeneous and bamboo-like NWs were imaged. In most of the cases, images like those shown in **Figure 2** are observed. The magnetic image in **Figure 2a** corresponds to the homogeneous diameter nanowire, where only a main dark and bright contrast is noticeable at the ends of the NW. That leads us to conclude its single domain configuration. However, the MFM image in **Figure 2b**, corresponding to the bamboo NW, displays the same dark and bright contrast at the ends plus additional, less intense, periodic contrasts along the wire. The comparison between magnetic and topographic images demonstrates the correlation between the periodic modulations in the diameter and the low-contrast MFM magnetic modulations. The MFM images thus unveil an essentially single domain configuration with a periodic local spin divergence induced at the sites of increased diameter.

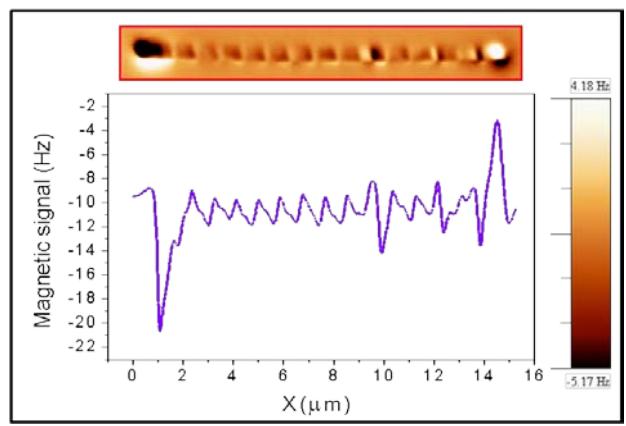

**Figure 3.** The figure shows an example of a nanowire with several strong black-white contrast in the positions of enlarged diameter.

Although the configuration shown in **Figure 2** is the most usual, we have also found a different behavior in a few cases, where high contrast (bright and dark) spots are identified in certain positions that match with enlarged diameter regions along the NW (see **Figure 3**). Such strong periodic contrast could be eventually interpreted as originated by the pining of domain walls at the diameter modulations. In order to obtain deeper insight into the magnetic configuration in all the cases and their evolution under applied magnetic field, an advanced variable field MFM technique has been applied.

### 4. 3D mode MFM

Magnetization reversal processes in both homogeneous and bamboo nanowires were investigated by using the so-called 3D modes[24]. Special attention was paid to the pinning process. In this MFM based mode, a profile along the main axis of the desired nanowire is repeatedly scanned while an *in-situ* external magnetic field is swept typically between +/- 45mT.

The in-plane magnetic field, parallel to the main axis of the NWs, is not high enough to reach magnetic saturation in modulated NWs, but sufficient to reverse their magnetization. Further details are supplied in the **Supporting Information 1**.

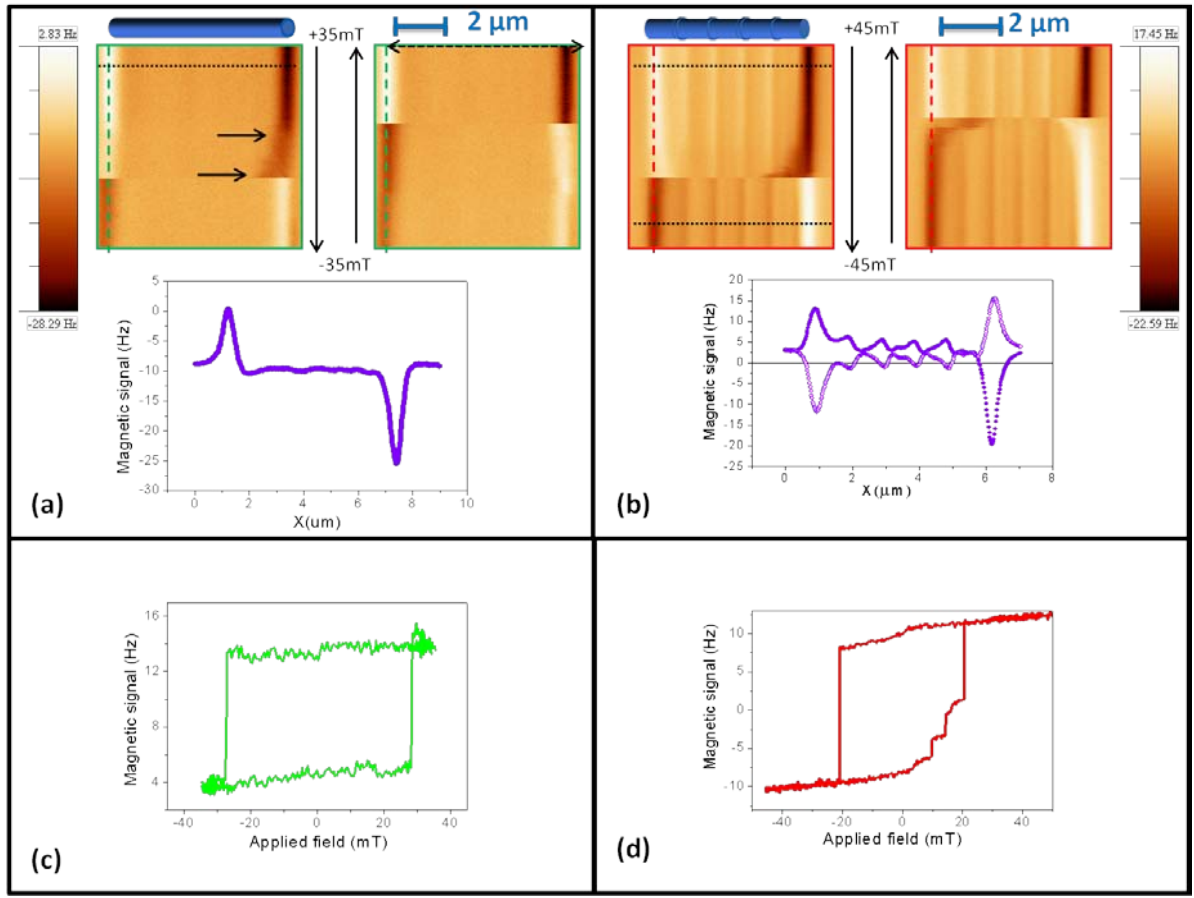

**Figure 4.** (a) 3D MFM based image of a nanowire with homogeneous diameter. Two critical fields have been marked with arrows: Profile measured along the dotted black line of the MFM image. (b) In the bamboo-like nanowires, a DW propagated towards the opposite end, jumping from one pinning site to the next. In these NWs, the saturation is not reached as shown in the profile measured along the dotted black line marked in **Figure 4b**. Based on the 3D images, non-conventional hysteresis loops of (c) homogeneous and (d) bamboo-like nanowires were depicted corresponding to dashed lines in (a) and (b).

**Figure 4a** and **Figure4b** show the evolution of the magnetic contrast under in situ magnetic fields for NWs with homogeneous and modulated diameter respectively. For high magnetic fields, the homogeneous NW is nearly saturated and thus the MFM signal exhibits the so called dipolar contrast (see the profile shown in **Figure 4a** measured along the dotted black line). As the field is swept, two critical fields are observed (marked with arrows in **Figure 4a**). As the field is decreased from its highest value, the contrast slightly reduces while remaining concentrated at the very end. We assume a continuous spin reorientation at very local scale at the end of the nanowire that gives rise to the formation of a kind of closure domain or vortex -as predicted for NWs with diameters beyond a few tens of nanometers[28]. In the second critical field, the magnetic contrast at the nanowire ends is suddenly reversed when the closure structure becomes energetically unstable, a domain wall is thus depined and propagates along the NW length resulting in a single large Barkhausen jump.

The 3D mode MFM images allow us to obtain non-conventional hysteresis loop by measuring the evolution of the contrast in one edge of the NW (along the dashed green line marked in **Figure 4a**). MFM data in **Figure 4a** and **Figure 4c** suggests a bistable behavior where either a positive or negative remanent configuration is possible. Notice that, in principle, DWs can nucleate simultaneously in both ends. However, the local geometry of each particular NW determines whether one or two DWs are involved in the magnetization reversal[29]. Slight differences in the critical fields between different nanowires are expected due to small morphological differences amongst them.

Unlike in homogeneous nanowires, 3D measurements corresponding to bamboo NWs outstandingly show the intermediate pinning of the DW in specific diameter modulation sites near the ends of the NW (see the right part of the 3D image corresponding to the branch from +45 to -45mT in **Figure 4b**). It should be noticed that the contrast at the enlarged diameter regions remains even at the highest applied field (see **Figure 4b**). At a critical field a DW depins and propagates along the NW. It is important to emphasize that at the diameter modulations the contrast is reversed as the DW passes through. In the following 3D image corresponding to the branch from -45 to +45mT, we observe even more clearly that the DW stops in two identified modulated sites giving rise to small jumps in the hysteresis loop until finally reaching the far end of the NW. Since the driven parameter, the magnetic field, is sweeping continuously, after the first depinning, the applied magnetic field is much higher that the critical field i.e. the domain wall speed increases[30,31].

Small variations in the DW propagation and asymmetries are measured in successive field cycling for the same nanowire which evidence how critical the data acquisition speed might be in these measurements. We must keep in mind that the scanning speed of the MFM is around 100 microns per second while the typical domain wall propagation velocity is 6 orders of magnitude higher.

The normalized hysteresis loops of the modulated NW -shown in **Figure 4d**- were obtained from the 3D MFM image in **Figure 4b** (measuring along the vertical dashed lines). However bamboo-like NWs present several intermediate configurations and therefore, the raw data from one of the nanowire ends is not representative of the whole NW. To obtain a conventional hysteresis loop a reconstruction is needed. Using these 3D modes, we have calculated an average hysteresis loop for the modulated NW that allows us to give an average coercive value of 25mT. See **Supporting information II**- for more details.

5. Domain Wall pinning

Coming back to the atypical behavior shown in **Figure 3**, we have used the 3D modes to determine whether or not there are pinned domain walls.

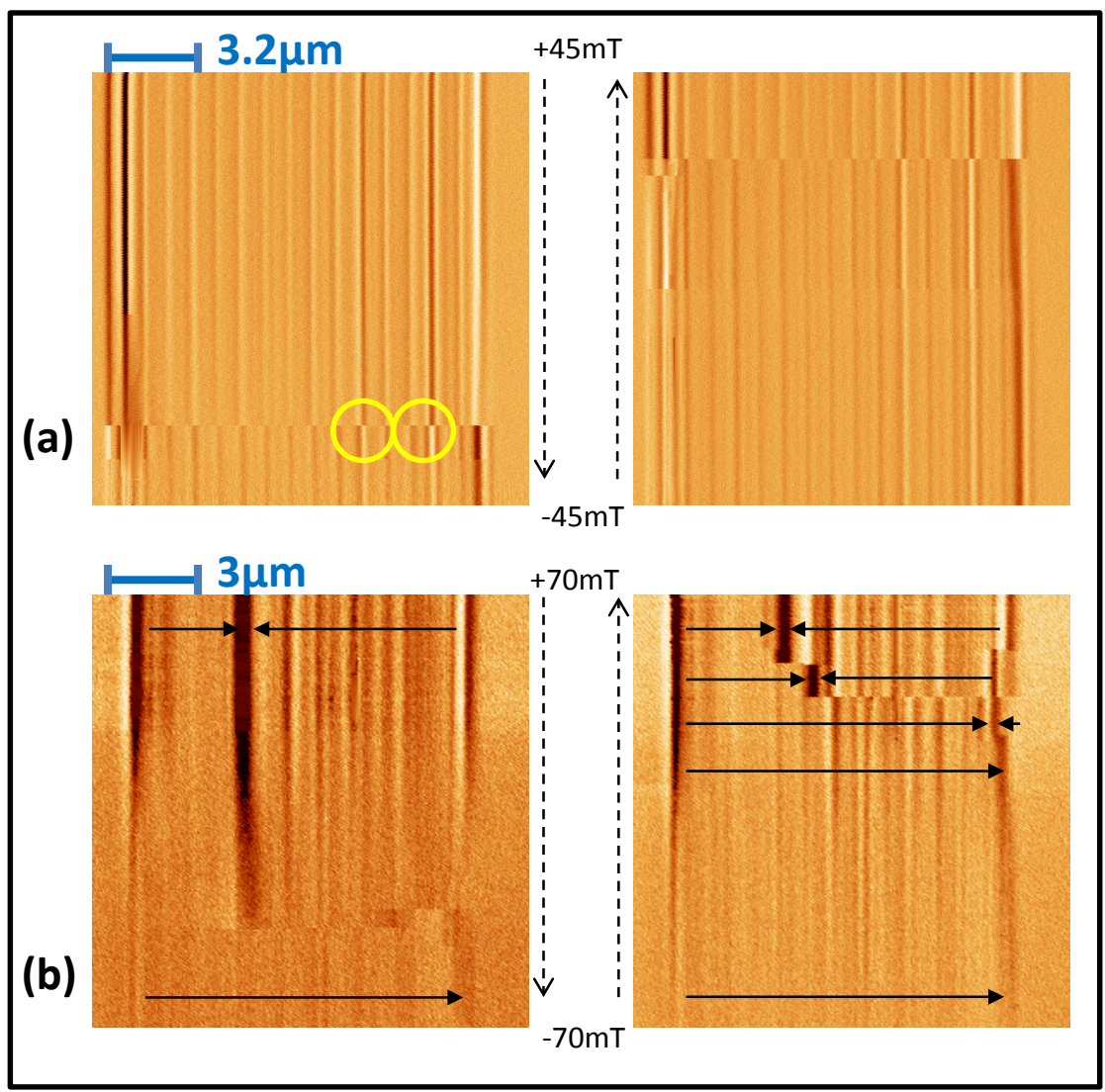

**Figure 5. (a)** 3D mode images of a NW showing strong contrast at some intermediate modulations. **(b)** 3D mode images corresponding to a NW with pinned domain walls. In these experiments the magnetic field varies between +/- 70mT. The black arrows indicate the direction of the magnetization.

As already shown in **Figure 3**, in some cases, strong black-white contrast is found, not only at the edges, but in the middle of the nanowire. Such features observed in remanence can be ascribed to the existence of domain walls pinned at these positions. However a deeper analysis of the 3D modes performed in these NW reveals that in most of the cases the contrast does not move towards the end of the nanowire as expected **(see Figure 5a,** marked in yellow). Moreover, similarly to data in **Figure 4b**, the contrast of these spots is reversed when the magnetization of the whole wire changes its direction as displayed in the marked regions of **Figure 5a**. This behavior cannot be explained with the existence of a pinned domain wall but with the existence of high stray field regions induced by defects.

A different behavior is exhibited by the NW shown in **Figure 5b.** In this particular case, where the 3D mode measurement has been performed starting from a negative remanent state, the field is switched between +/- 70mT . Again strong contrast at the middle positions of the NW is observed at the remanence. A higher field of around 50 mT is now required to observe reversal of local contrasts at modulations. Moreover, we observe that several local reversals are detected at different positions of the NW. That would confirm: i) the presence of few DWs pinned in some modulations; ii) the existence of several domains at the remanent state with opposite magnetization; iii) the motion of a single DW along the NW (see the arrows in the MFM image corresponding to the branch from -70mT to +70mT). It is inferred from those results that although in general there are no domain walls pinned in the middle of the NWs, some bamboo-like nanowires present pinned domain walls that can be controlled by an external magnetic field. As expected, these nanowires are magnetically harder than the others.

## 6. Micromagnetic simulations

As mentioned above, many authors propose a vortex (or a system of vortices) as the expected configuration for this kind of nanowires. MFM data in combination with micromagnetic simulations shed light on the spin configuration of the bamboo-like nanowires.

As shown in the micromagnetic simulations in **Figure 6**, the spin configuration is a combination of a vortex -where the spins follow the enlarged diameter shape- plus a configuration with the core spins aligned with the cylinder axis. The simulated spin configuration displays two equal vortices appearing at both edges of the NW (i) and (iii). However, at the position with enlarged diameter (ii) a pseudo-vortex with a large core is expected. Notice that the MFM is hardly sensitive to this pseudo-vortex due to the lack of stray field and then cannot be the origin of the MFM contrast. Nevertheless, the bright and dark contrasts are induced by the previous and subsequent high radial stray field regions. The 3D mode images is agreement with this configuration, when the magnetization along the nanowires is reversed, the contrast on the modulation also inverts.

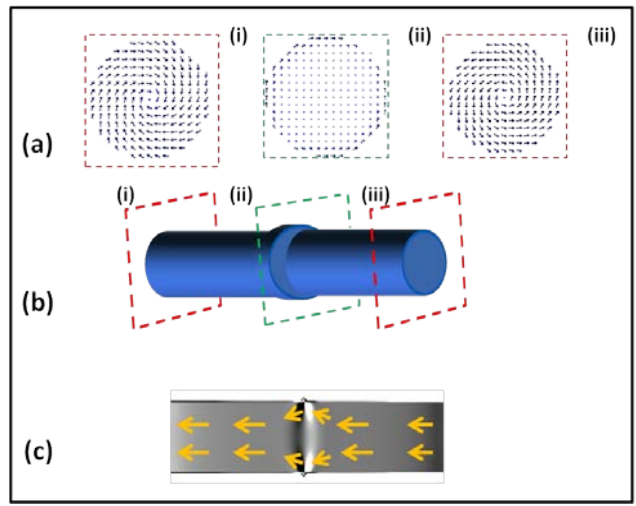

**Figure 6.** Micromagnetic simulations of the spin configuration of a modulated NW calculated by OOMMF (a) Different vortex configuration appear along the NW (i) , (ii) and (iii). Notice that the chirality of the pseudo-vortex in (ii) is opposite to the vortices generated at the edges. (c) In a section along the wire, magnetization divergence plus some arrows indicating the magnetization direction are depicted.

The micromagnetic simulations are in good agreement with the MFM data corresponding to the standard bamboo-like NWs without pinned domain walls. Thus, it is demonstrated that the pinning at intermediate sites is a metastable configuration. Nevertheless, in some cases, we have achieved to create domain walls and control them by an external magnetic field. The origin of this atypical behavior can be the high density of defects at the enlarged diameter regions that presents decreased pinning energy.

## 7. Conclusions

The aim of this research is to study the pinning effect induced by the diameter modulation and to determine the magnetization reversal process in a single bamboo-like nanowire by high sensitivity Magnetic Force Microscopy.

Two different kinds of ferromagnetic nanowires -grown by electrodeposition in alumina templates- were studied: homogeneous diameter NWs and bamboo-like NWs. In the demagnetized state, the homogeneous NWs present single domain configuration while the magnetization in the modulated NWs exhibit strong stray fields at the enlarged diameter positions as displayed in the magnetic images. Several homogeneous and diameter modulated nanowires were measured for statistics and, on average, both kinds of NWs present coercive fields around 20mT.

Nevertheless, the two samples show different magnetization reversal processes. The MFM data reveals that while the homogeneous NWs present an abrupt magnetization reversal through the depinning and propagation of a single DW, a metastable intermediate pinning has been measured in most of the the bamboo-like NWs. It is worth mentioning that the 3D mode MFM imaging allows us to distinguish two different magnetic configurations in bamboo-like NWs that can be misinterpreted by using standard MFM images: (i) regions of high stray field and (ii) real domain walls. The nanowires with pinned domain walls present higher coercive fields, around 50mT, as measured from the 3D MFM modes.

In conclusion, these results demonstrate that the bamboo-like NW are promising structures in future spintronic devices, and further investigations should adressed to gain better control over the DW pinning.

## 8. Experimental Section

*Sample preparation*

The bamboo-like nanowires were produced using the self-assembled pores of alumina templates obtained by pulsed hard anodization in oxalic aqueous solution (0.3M) containing 5 vol.% ethanol at a constant temperature of 0°C[32,.33]. In a first step a constant voltage of 80 V was applied for 900s to produce a protective aluminum oxide layer at the surface of the disc which avoids breaking or burning effects during subsequent hard-pulse anodization[34,35]. Afterwards, the voltage is steadily increased (0.08V/s) to 140V and kept constant for 600s, which ensures the parallel alignment of the nanochannels. The modulated nanopores were produced by periodically applying pulses of 140V and 80V for 30 and 10 s, respectively. The pulses were repeated 30 times to guarantee a total length of the modulated pores of few tens of microns. The wires were grown into the alumina pores by electrodeposition from a sulfate-based electrolyte [36]. The resulting periodically modulated pores are formed by 800 nm long segments, 150 nm in diameter, separated by much shorter segments, few tens of nanometer long, 170 nm in diameter, forming a bamboo-like structure as observed in Fig.1. The center-to-center inter-pore distance is kept constant at 320 nm.

*Magnetic Force Microscopy measurements*

A scanning force microscope from Nanotec Electronica has been used to perform all the measurements, together with Nanosensors PPP-MFMR microchips. Amplitude modulation method was performed and the phase-locked loop was enabled to track the resonance frequency of the oscillating cantilever.

*Micromagnetic Simulations*

A 780 nm long cylinder has been simulated, with a diameter of 150nm. To study the effect of the modulation, a broadening in the diameter has been included in the middle of the nanowire, reaching a maximum diameter of 170 nm. Regarding the characteristics of the material, it was assumed that FeCoCu lacks of crystalline anisotropy ($K_1=K_2=0$) and magnetization and exchange coupling constant were assumed to be $M_s= 14 \times 10^5$ A/m and $A= 10.7 \times 10^{-12}$ J/m respectively. The cubic cell size was chosen 2.5nm, to be below the exchange length which is approximately 3nm for this material.

**Acknowledgements**
This work was supported by Spanish MINECO (CSD2010-00024 and MAT2013-48054-C2).

Received: ((will be filled in by the editorial staff))
Revised: ((will be filled in by the editorial staff))
Published online: ((will be filled in by the editorial staff))

Supporting Information

1- **3D modes**

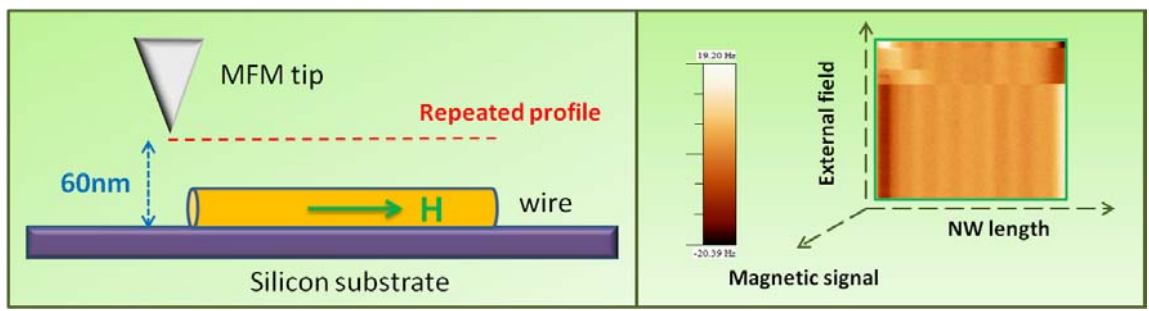

**Figure S1.** (a) The followed procedure to obtain data with the 3D modes is illustrated. (b) Typical image obtained by 3D modes.

Non-conventional MFM images (**Figure 4a** and **Figure 4b**) are obtained as follows: the tip is placed onto the desired nanowire and the scans are performed along the NW length at a typical distance of 60nm, as shown in **Figure S1a**. This scan line is repeated whereas the external applied magnetic field is swept between the maximum and the minimum of the magnetic field applied parallel to the nanowire axis.

As a result, in the 3D mode image (**Figure S1b**), X axis represents the length of the scanned line, Y the external applied magnetic field and Z the MFM signal for every position on the nanowire and every magnetic field value. The two images presented for the two samples (**Figure 4a** and **Figure 4b**), complete the full minor hysteresis loop, the first image corresponds to a field sweep from +35 to -35mT (+45 to -45mT in the case of **Figure 4b**), while the second corresponds to a field sweep beginning from -35mT to +35mT (-45mT to +45mT in the case of **Figure 4b**).

## 2- Reconstruction of the hysteresis loops

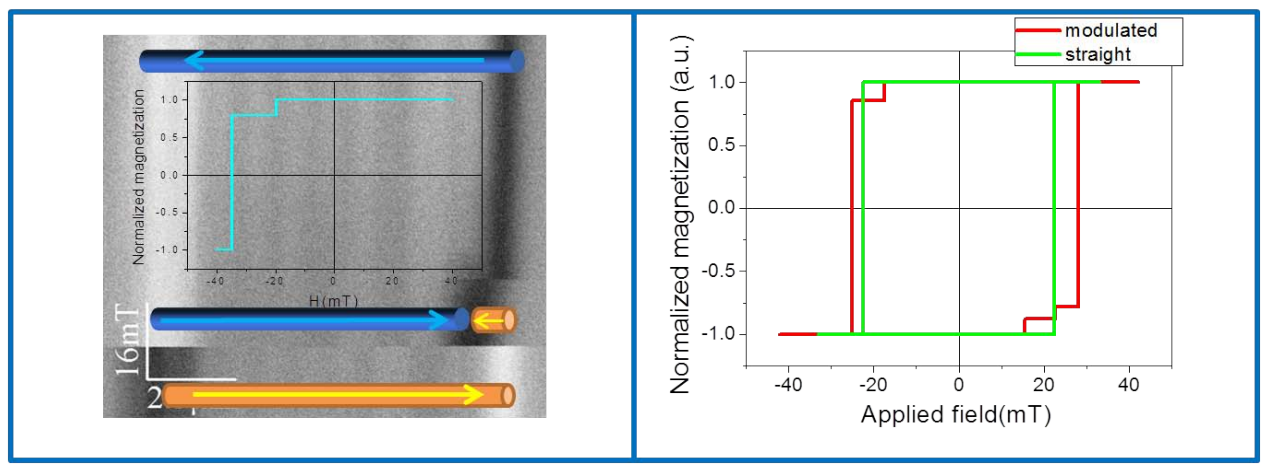

**Figure S2.** (a) The normalized magnetization is calculated for each configuration using geometrical considerations. (b) Reconstructed hysteresis loops for both kinds of nanowires.

We have evaluated the changes of the magnetic moment along the axis associated to the jumps as shown in the **Figure S2**. In this simple model the initial and final magnetization values are considered -1 and 1 respectively, and the values from the configurations in between are calculated from purely geometrical considerations.

The hysteresis loops obtained when we plot the profiles obtained from the MFM data are similar to the conventional hysteresis loops which quantify the magnetization in the NW axis as a function of the applied field. Nevertheless, we have to keep in mind that MFM is sensitive to the stray field of the sample, in the out of plane direction. This means that, especially in the case of the diameter modulated NWs, although the hysteresis curves might seem similar to the conventional, the size of the Barkhausen jumps is not real. Further calculus need to be performed, since a single profile at the edge gives only information of the local changes in the magnetization.

Supporting Information is available from the Wiley Online Library or from the author.